# Nanoscale

# Solute particle near a nanopore: influence of size and surface properties on the solvent-mediated forces†

Julien Lam  and James F. Lutsko  *

Nanoscopic pores are used in various systems to attract nanoparticles. In general the behaviour is a result of two types of interactions: the material specific affinity and the solvent-mediated influence also called the depletion force. The latter is more universal but also much more complex to understand since it requires modeling both the nanoparticle and the solvent. Here, we employed classical density functional theory to determine the forces acting on a nanoparticle near a nanoscopic pore as a function of its hydrophobicity and its size. A simple capillary model is constructed to predict those depletion forces for various surface properties. For a nanoscopic pore, complexity arises from both the specific geometry and the fact that hydrophobic pores are not necessarily filled with liquid. Taking all of these effects into account and including electrostatic effects, we establish a phase diagram describing the entrance and the rejection of the nanoparticle from the pore.

## 1 Introduction

Determining forces acting on a nanoparticle in the vicinity of a nanoscopic pore is of great scientific and industrial importance for a wide range of systems such as filtering membranes for water purification[1–6] and porous solids made of cylindrical cavities employed as nucleants for selective crystallization.[7–9] In biophysics, nanopores are used for protein unfolding,[10–15] DNA sequencing[16–21] and protein detection.[22–26] In all these cases, controlling the solute behavior close to the pore entrance plays a major role in the success of the desired application and contributes to the ultimate goal of engineering systems to have specific properties.

When studying this type of system, one major difficulty consists of modeling the depletion force accounting for the indirect role of solvent molecules. Depletion forces are observed in numerous fundamental processes such as colloidal self-assembly,[27–30] protein stabilization in helix[31,32] and the effective attraction between hydrophobic surfaces.[33,34]

Center for Nonlinear Phenomena and Complex Systems, Code Postal 231, Université Libre de Bruxelles, Boulevard du Triomphe, 1050 Brussels, Belgium.
E-mail: jlutsko@ulb.ac.be

†Electronic supplementary information (ESI) available: (1) Computation of the contact angle relationship with the energy parameter ε, (2) density profile, (3) geometrical expressions used to compute the depletion forces using a capillary model, (4) model to identify conditions at which the cylindrical pore is filled/empty, (5) comparison of density profiles for small hydrophilic pores and (6) nanoparticle inclusion for empty and filled pores (7) role of the nanoparticle hydrophobicity with a hydrophobic wall. See DOI: 10.1039/C7NR07218J

Microscopically, depletion forces result from two effects: (a) at very short separations, solvent molecules are excluded from the gap between the macroscopic objects (particles, molecular chains and flat walls) thus generating a force; and (b) the formation of low-density phases in the gap between the objects due to capillary effects. The first contribution to the depletion force only depends on the geometry and the surface properties of the macroscopic objects whereas the second depends as well on the interaction between the surfaces and the solvent. These solvent-mediated interactions were thoroughly reviewed by Chandler[35] and also by Berne et al.[36] In general, many different model liquid solvents have been studied including hard sphere,[27,37–44] Lennard-Jones potentials[45–47] and also single point charge water.[35,48–51] In addition, various specific geometries have been discussed in the literature such as biomolecules,[52] two big spheres,[27,37] two planar surfaces[40,41,45] and a big sphere and a planar surface.[37–39]

However, despite its practical importance, the interaction between a nanoparticle and a nanoscopic pore has received less attention due to its inherent complexity. Previous works thoroughly studied the behavior of a large hard sphere in the vicinity of a rigid non-interacting wall and thus only focussed on entropic effects.[42–44] In addition, the time scale for the nanoparticle entrance inside a pore was discussed recently for a spherical pore of a fixed size and hydrophobicity.[53] In this paper, we aim at studying how the interplay between hydrophobicity and aspect ratio of the pore can induce the nanoparticle entrance or rejection. We model the liquid using a simple Lennard-Jones potential so as to describe the dominant effects of excluded volume and nontrivial energetics giving rise to wetting behaviour. Finite-temperature density functional theory[54,55] (DFT) was employed to compute the equilibrium state including the fully (non-symmetry constrained) three dimensional density profile along with the free energy. We applied the model first to a nanoparticle interacting with a flat surface and inside an infinite cylinder. From these calculations, a capillary model is built to understand the mechanisms generating the depletion forces. Then, we focused on a system made of a solute in the vicinity of a nanoscopic pore for empty, filled and partially filled pores and we determined under which conditions the nanoparticle enters the pore. Moreover, we introduced functionalization of the surfaces modeled as electrostatic forces and show that when these are sufficiently high, the nanoparticle entrance can be activated even in conditions where the depletion forces would dictate otherwise. Finally, the influence of the nanoparticle hydrophobicity was studied in the last section. Our work contributes to the general understanding of solute infiltration within nanoscopic pores by providing a qualitative picture guiding the design of pores in terms of size, solvophobicity and chemical functionalization. In addition, the quantitative figures obtained in this work provide a basis for a coarse-graining of the solute–pore interactions. This should ultimately lead to simplifications of the multi-scale problems made of solvent/solute/pore.

## 2 Theory

### 2.1 Density functional theory

Our systems can be divided into three elements: the walls, the nanoparticle and the fluid. The walls and the nanoparticle are static and play the role of external potentials acting on the fluid. The local density of the fluid and the free energy of the entire system are calculated using classical Density Functional Theory. First, an energy functional is constructed as

$$\Omega[\rho] = F_{id}[\rho] + F_{HS}[\rho] + F_{att}[\rho] \\ + \int \rho(\mathbf{r})(\phi_{ext}(\mathbf{r}) - \mu)\mathrm{d}\mathbf{r} \tag{1}$$

where $\rho(\mathbf{r})$ is the local number density of the fluid and $\mu$ is the chemical potential. This functional is minimized with respect to the density to obtain the equilibrium, non-uniform density of the fluid. Evaluating $\Omega[\rho]$ at this minimizing density distribution gives the grand-canonical free energy of the system.[54,55]

The ideal gas contribution is

$$F_{id}[\rho] = k_B T \int (\rho(\mathbf{r}) \ln \rho(\mathbf{r}) - \rho(\mathbf{r}))\mathrm{d}\mathbf{r}. \tag{2}$$

The molecules of the fluid interact *via* a pair potential, $v(r)$, which we take to be a Lennard-Jones interaction that has been truncated at $r = r_c$ and shifted giving

$$v(r) = v_{LJ}(r) - v_{LJ}(r_c), \quad r < r_c \tag{3}$$

and zero for $r > r_c$. This potential is then separated into a long-ranged attraction, $w(r)$ and a short-ranged repulsion, $v_{rep} = v(r)$

$- w(r)$ using the Weeks–Chandler–Anderson scheme whereby $w(r) = v(r_{min})$ for $r < r_{min}$, where $r_{min}$ is the position of the minimum of the potential, and $w(r) = v(r)$ for $r > r_{min}$. The repulsive part is used to calculate a temperature-dependent effective hard-sphere radius *via* the Barker–Henderson prescription,

$$d_{eff}(T) = \int_0^{r_{min}} \exp(-\beta v_{rep}(r))\mathrm{d}r, \tag{4}$$

which is used in turn to specify the hard-sphere contribution, $F_{HS}[\rho]$, as the density functional for a system of hard spheres have diameter $d_{eff}(T)$. For this, we have used the White Bear functional with tensorial densities.[56] The contribution of the attractive part of the potential is calculated using the mean-field expression,

$$F_{att}[\rho] = \frac{1}{2}\int \rho(\mathbf{r}_1)\rho(\mathbf{r}_2)v_{att}(\mathbf{r}_{12})\mathrm{d}\mathbf{r}_1\mathrm{d}\mathbf{r}_2. \tag{5}$$

Finally, the external potential $\phi_{ext}(\mathbf{r})$ at any given point is the sum of the contributions from all of the walls and the nanoparticle.

We perform the calculations on a three dimensional grid. The fluid density is specified on the grid points and the free energy is minimized with respect to these values using conjugate gradients. The hard-sphere contributions require performing spherical integrals of the density and this is done using trilinear interpolation of the density and spherical t-designs.[57] For some of our applications, we minimize at constant particle number, which is the integral of the density over all space, rather than at constant chemical potential. Technically, there are additional corrections that should be evaluated in this case[58] but in practice it is known that these tend to be negligible for systems of the size considered here.

To carry out the numerical calculations, the density is discretized on a Cartesian lattice and the entire system is subject to periodic boundaries. Care was taken to extend the systems sufficiently in each direction so as to avoid self-interactions, *etc.* All calculations were carried out with 5 lattice points per unit of $\sigma$ which is sufficiently small to perform calculations for large simulation boxes while retrieving quantitative agreement with finer discretizations.

### 2.2 Model

For the liquid/liquid interaction, the Lennard-Jones potential is parametrized by the length scale denoted $\sigma$ and the energy parameter denoted $\varepsilon_{liq/liq}$. The cutoff is chosen equal to $r_c = 3\sigma$. We worked at a temperature of $k_B T = 0.8\varepsilon_{liq/liq}$ which is located between the triple point and the critical temperature[59] but we do not believe that this choice affects physics discussed below. In addition, the chemical potential is chosen higher than the one at liquid–vapor coexistence denoted, $\mu_{coex}$. Indeed, at room temperature and at atmospheric pressure, liquid water does not coexist with its vapor state. The pressure at coexistence is roughly $P_{coex} = 1/20 P_{atm}$ so in the DFT calculations, the chemical potential is chosen to reproduce this ratio of the pressure to its value at coexistence resulting in a supersaturation $\Delta\mu \equiv \mu - \mu_{coex} = 0.27 k_B T$.

Solids are often represented by effective potentials such as the Steele potential[60,61] and related models[62] which are derived for simple geometries like planar walls. Here, we will address the complex geometry of a cylindrical pore so we have modeled our solids within an atomistic framework. In practice, the solid is represented by fixed particles disposed in a face-centered cubic (FCC) lattice with the (100) surface exposed to the fluid. The initial solid density is $\rho = 1.09\sigma^{-3}$ which is the equilibrium, zero-temperature density for an FCC Lennard-Jones solid.[63] For each calculation, the simulation box lengths are chosen to be commensurate with the initial lattice spacing of the wall and then readjusted to also be commensurate with the length scale of the DFT discretization used for the calculation. The lattice spacing is then modified to be a multiple of the final simulation box lengths which changes the solid density by a negligible amount (no more than 1%). The same truncated and shifted Lennard-Jones potential is used to model solid–liquid interactions. For simplicity and to reduce the number of parameters the length parameters for both interactions are taken to be equal. With such a model, the wettability represented by the contact angle, $\theta$, is directly related to the ratio between the two Lennard-Jones energy parameters: $\varepsilon^* \equiv \varepsilon_{wall/liq}/\varepsilon_{liq/liq}$. Details on the contact angle derivation can be found in the ESI, section 1.1.† In practice, for large values of $\varepsilon^*$, liquid molecules are more attracted to the solid than to themselves and the solid is hydrophilic.

Initially, we take the nanoparticle to be a hard sphere particle of radius $R_n$ thus avoiding considerations regarding the chemical nature of the solute to focus only on the depletion forces. Later, we discuss the effect of direct interactions between the nanoparticles and the walls. To fix the relevant length scales and to explain our choices of nanoparticle sizes, we note that a water molecule has a typical effective size (diameter) of about 0.275 nm while typical globular proteins such as Lysozyme have diameters of 3 or 4 nm so that the smallest nanoparticles of interest here have a diameter on the order of 10 water molecules and, hence, a radius of about $R_n = 5\sigma$.

## 3  Results

### 3.1  Depletion potential in the case of a flat wall and an infinite cylinder

*Flat wall.* The solid is made of 6 layers of an FCC crystal which is sufficient given the cutoff of the potential and the entire simulation domain is of size $61.8\sigma \times 61.8\sigma \times 36.0\sigma$. The nanoparticle is placed at various distances from the wall and both the liquid density profile and free energy are computed by minimizing the free energy functional. Fig. 1 shows typical density profiles obtained with this geometry. Near the nanoparticle the liquid arranges itself into spherical shells due to the excluded volume of the liquid molecules. Similarly, layering of the liquid near the wall is also observed and when the walls are hydrophilic, the liquid begins to adapt to the lattice structure of the solid wall.

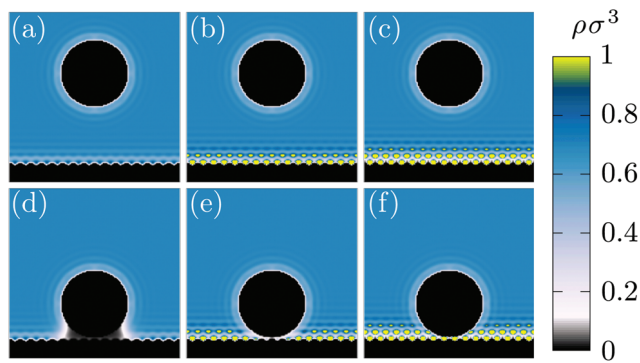

**Fig. 1** Liquid density profiles for a nanoparticle, $R_n = 5\sigma$, near a flat wall for various hydrophobicities: (a, d) $\varepsilon^* = 0.1$ i.e. $\theta = 171°$, (b, e) $\varepsilon^* = 0.3$ i.e. $\theta = 95°$ and (c, f) $\varepsilon^* = 0.5$ i.e. $\theta = 0°$. The distance between the nanoparticle center and the wall is equal to $8\sigma + R_n$ for (a, b, c) and $R_n$ for (d, e, f). Each image is of size $25\sigma \times 25\sigma$ [see also ESI, section 2†].

The free energy as a function of the nanoparticle/wall distance, denoted $h$, is shown in Fig. 2 for different contact angles and for two solute diameters. Far from the wall ($h > R_n + 5\sigma$), the free energy is constant and there is no effective interaction between the nanoparticle and the wall. Closer to the wall, the behaviour of the free energy depends on the degree of hydrophobicity. Qualitatively, two effects are at work. First, it is energetically costly for fluid to be in contact with a hydrophobic wall – alternatively, one can say that the fluid–wall surface tension increases with the degree of hydrophobicity. In this case, the presence of a nanoparticle near the wall allows for the formation of a region of vapor-like density which has a lower wall surface tension and therefore lowers the total free energy of the system. Second, the usual depletion force (effective attraction due to exclusion of the solvent) gives rise to an attractive interaction even for a purely hard-core interaction between the wall and the fluid.[45] For this reason, the effective force on the nanoparticle does not become negative at the wetting transition (where the fluid–wall interaction becomes favored with respect to the vapor) but rather at higher degrees of hydrophilicity in which the fluid has an effectively negative wall surface tension thus overwhelming all other effects. Finally, the forces are higher when the nanoparticle is larger because more solvent molecules are influenced by the nanoparticle presence.

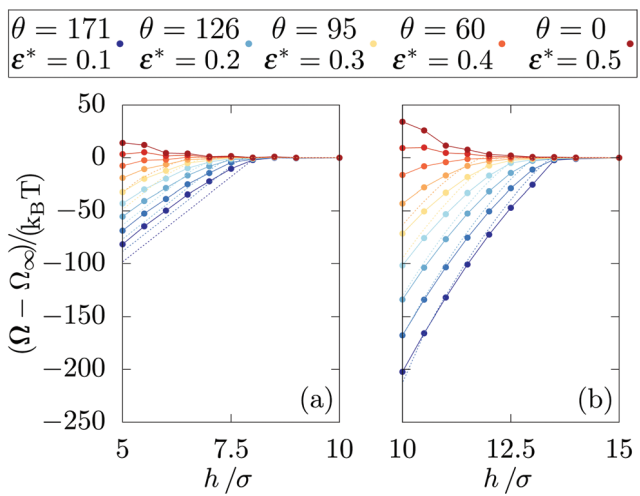

**Fig. 2** Depletion potential (excess grand potential) for a nanoparticle of different radii (a) $R_n = 5\sigma$ and (b) $R_n = 10\sigma$ as a function of the distance between the nanoparticle and the wall. Different degrees of hydrophobicity were computed. Colors for contact angle $\theta$ equal to 144°, 110°, 79° and 36° are not showed in the legend. Dotted lines correspond to the capillary model results.

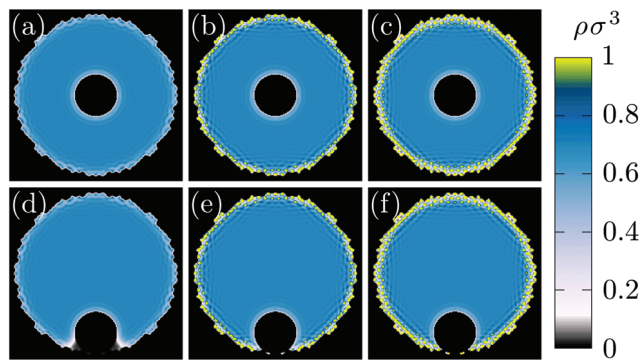

**Fig. 3** Liquid density profiles for a nanoparticle, $R_n = 5\sigma$, inside a cylinder of diameter, $D_{cyl} = 37.5\sigma$ for various hydrophobicity: (a,d) $\varepsilon^* = 0.1$ i.e. $\theta = 171°$, (b, e) $\varepsilon^* = 0.3$ i.e. $\theta = 95°$ and (c, f) $\varepsilon^* = 0.5$ i.e. $\theta = 0°$. The distance between the nanoparticle center and the cylinder wall is equal to $D_{cyl}/2$ for (a, b, c) and $R_n$ for (d, e, f). Each image is of size $40\sigma \times 40\sigma$ [see also ESI, section 2†].

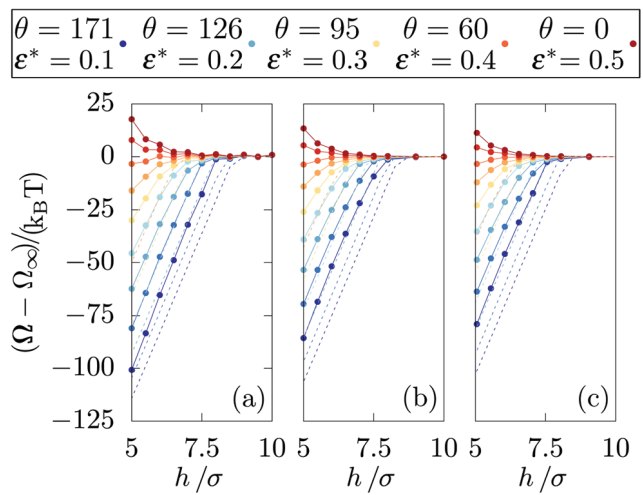

**Fig. 4** Depletion potential (excess grand potential) for a nanoparticle inside an infinite cylinder as a function of the distance between the nanoparticle and the wall under various degrees of hydrophobicity. The cylinder diameter is equal to (a) $D_{cyl} = 25\sigma$, (b) $D_{cyl} = 37.5\sigma$ and (c) $D_{cyl} = 50\sigma$. Points and plain lines correspond to the DFT results. Colors for $\theta$ equal to 144°, 110°, 79° and 36° are not shown in the legend. Dotted lines correspond to the capillary model results.

*Infinite cylinder.* In this case, the system is made of a large FCC crystal out of which all particles within the given diameter ($D_{cyl}$) are removed. The length of the cylinder is equal to $31\sigma$ and the periodic boundaries imply that it is effectively infinite in length. For the radial dimensions, the crystal size is chosen so that it has a width greater than $r_c$. Both the density and free energy are computed as a function of the distance from the nanoparticle to the cylindrical wall. Fig. 3 shows typical density profiles obtained with the described geometry.

As shown in Fig. 4, similarly to the case of a flat wall, increasing the hydrophobicity leads to an increase of the solute attraction. In addition, the forces are higher inside smaller cylinders since layering is more present.

*Capillary model.* For both a flat wall and a cylinder, two regimes are observed: (i) far from the wall, the free energy is constant, (ii) close to the wall, the free energy decreases almost linearly. As evidenced by Fig. 1d and 3d, this abrupt transition results from the formation of a low-density gas-like phase between the wall and the nanoparticle. This suggests that a simple capillary model can be used to estimate the excess free energy when the nanoparticle approaches the wall. We approximate the gaseous volume as a cylinder stretching from the wall to the nanoparticle with a radius of $r_0$ which is, in general, not equal to $R_n$. We assume the low-density fluid has the pro-perties of the bulk gas phase and introduce $\gamma_{n,l/g}$, $\gamma_{w,l/g}$ and $\omega_{l/g}$ as respectively the nanoparticle–liquid/gas surface tensions, the wall–liquid/gas surface tensions and the bulk liquid/gas grand potential per unit of volume. Accordingly, the excess grand potential, $\Omega(h) - \Omega(h \to \infty)$, is given by:

$$\Omega(h) - \Omega(h \to \infty) = (\omega_g - \omega_l)V_{cyl} + \gamma_{l,g}S_{cyl} + (\gamma_{w,g} - \gamma_{w,l})S_{wall} + (\gamma_{n,g} - \gamma_{n,l})S_{nano}$$

(6)

where all the constants are determined using DFT in the grand canonical ensemble and the geometrical factors are expressed in the ESI, section 3.† Note that, in this model, the dependence on surface properties is dual: (i) $\gamma_{w,g} - \gamma_{w,l}$ explicitly varies with the contact angle, (ii) $r_0$, because it measures the size of the gaseous region, also depends implicitly on surface properties.

As shown in Fig. 2 and 4, the DFT results are in good qualitative and indeed semiquantitative agreement with predictions from the capillary model for both geometries. The slight shift can be due to the gaseous region approximated as a purely cylindrical geometry thus neglecting edge effects close to the nanoparticle and the wall. Similar results were also obtained using single point charge model for the liquid and with different geometries[35,36,51] thus confirming the ability of our elementary model to capture most physical features underlying the hydrophobic interactions.

## 3.2 Depletion potential in the case of a cylindrical pore

Nanoscopic pores are built using an FCC crystal out of which atoms located inside a cylinder of diameter denoted $D_{cyl}$ and of height $H = 25\sigma$ are removed. The distinction from the pre-viously considered infinite cylinder is that the pore is terminated at one end by crystal an is open at the other end to the bulk fluid [see Fig. 6]. The height of the crystal is chosen to leave few layers of atoms below the pore and we fixed the width of the crystal equal to $D_{cyl} + 26\sigma$ which is sufficient for the pore sizes considered here. Above the wall, the system is loaded with a $25\sigma$ liquid layer [see Fig. 6].

*Liquid filling of the pore.* When changing the pore size, the wall hydro-affinity and also the degree of supersaturation, pores can be either empty or filled with liquid [see Fig. 6]. Furthermore, in many cases, both states are possible with one being metastable with respect to the other. Therefore, before computing the depletion forces, we will investigate which state is the most energetically favorable for the thermodynamic conditions to be considered. First, the density and free energy are computed starting from a fluid-filled pore, generally resulting in a fully filled pore after minimization of the free energy. The calculation is then repeated starting from an empty pore which can result in either a partially-filled pore with a meniscus or a fully filled pore. As a comparison, a capillary model is also used to determine the free energy of filled pores and of partially-filled pores with a meniscus [see ESI, section 4†]. Comparison of the free energies resulting from the two calculations allows for a determination of the thermodynamically favored state. At a supersaturation of $\Delta\mu = 0.27k_BT$, the pore is almost always filled with liquid [see Fig. 5a]. We therefore also studied the case of $\Delta\mu = 0.10k_BT$ for which, because of the lower pressure, partially-filled pores are more common. The phase diagrams show that the presence of a meniscus is favored by small hydrophobic pores and at low supersaturation. Finally, the agreement between the DFT and capillary-model calculations demonstrates that the latter captures the main physical features of the pore liquid filling.

*Depletion potential.* Then, calculations are run for a nanoparticle $R_n = 5\sigma$ at various positions and for different pore pro-perties ($D_{cyl}$, $\varepsilon^*$). The aim is to measure the grand potential and determine whether or not the nanoparticle is more likely to enter the pore.

Fig. 6 shows the grand potential for various positions of the nanoparticle at different pore conditions. In general, free energy is minimum when the nanoparticle is at the bottom of the pore for two different reasons. On the one hand, compared to when the pore is empty and the nanoparticle is in the bulk fluid above, inserting the nanoparticle inside the pore effectively transfers liquid molecules to the bulk region. Thus, the free energy decreases since the system is supersaturated and so favors the bulk liquid state. On the other hand, when the pore is filled, the cylindrical depletion forces are larger than those of a flat wall (as shown in Fig. 4). Consequently, when the nanoparticle manages to enter the pore it is more likely to stay inside.

However, when it is outside the pore, the nanoparticle does not "know" that the free energy will be lower at the bottom of the pore: whether or not it enters will depend on the free energy gradients near the pore entrance. Thus, we compare the free energy at the pore entrance (radial position $r < D_{cyl}/2$) with that far from it ($r = D_{cyl}/2 + 10\sigma$). These two free energies are respectively denoted $\Omega_{in}$ and $\Omega_{out}$ and indicated in Fig. 6. When $\Delta\Omega \equiv \Omega_{in} - \Omega_{out} > 0$, the nanoparticle is more likely to stay close to the wall rather than entering inside the pore. Fig. 7 shows $\Delta\Omega$ when changing the supersaturation, the pore size and its degree of hydrophobicity.

In summary, for pores much larger than the nanoparticle ($D_{cyl} \geq 2R_n = 10\sigma$), the particle is only favored to enter the pore at sufficiently high hydrophobicity. For pores with diameters less than twice that of the nanoparticle, entrance is increasingly favored with decreasing pore size until the particle prefers the pore even near the wetting transition. Finally, for pores smaller than the nanoparticle ($D_{cyl} = 7.5\sigma$), hydrophilic walls provokes the nanoparticle rejection. There, the pore are filled with highly structured liquid [see ESI, section 5†] and the nanoparticle presence at the pore entrance generates frustration. When the walls are sufficiently hydrophobic, the pore is empty thus avoiding frustration. These trends are stronger at higher chemical potential (*i.e.*, higher liquid density and pressure), probably due to stronger frustration of the fluid packing near the walls and nanoparticle. Surprisingly, at $D_{cyl} = 7.5\sigma$ and $\theta = 171°$, the nanoparticle is also rejected. There, the pore is empty and in general, increasing the hydrophobicity for an empty pore leads to an increase of the flat wall attraction while keeping the pore attraction constant [see ESI, section 6†].

### 3.3 Functionalization of the surfaces: the role of electrostatic effects

In addition to depletion forces, the nanoparticle can also interact with nanopores by means of electrostatic effects. The total free energy for a system with an explicit wall–nanoparticle interaction potential $V(\mathbf{r})$ and the system without such a potential is simply that $\Omega = \Omega_{V=0} + V(\mathbf{R})$ where $\mathbf{R}$ is the position of the center of the nanoparticle. Thus, our previous results can now be used to study the effect of such additional forces with little supplementary computation.

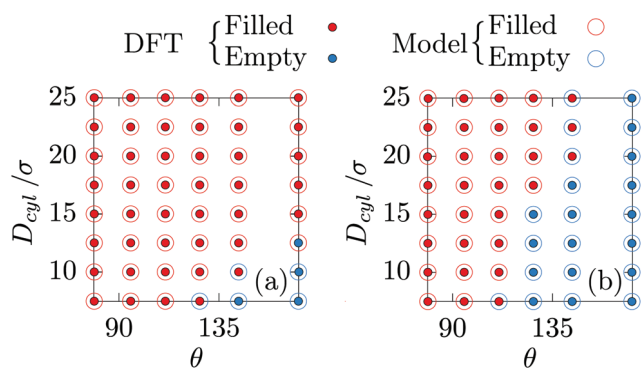

**Fig. 5** Phase diagram showing conditions for pore filling at two different supersaturations (a) $\Delta\mu = 0.27k_BT$ and (b) $\Delta\mu = 0.10k_BT$ as a function of the pore diameter $D_{cyl}$ and its hydrophobicity characterized by the contact angle $\theta$. Red and blue colors show conditions for which the most stable state is respectively a filled and empty pore. Filled (●) and open (○) circles represent respectively results from DFT calculations and the capillary model.

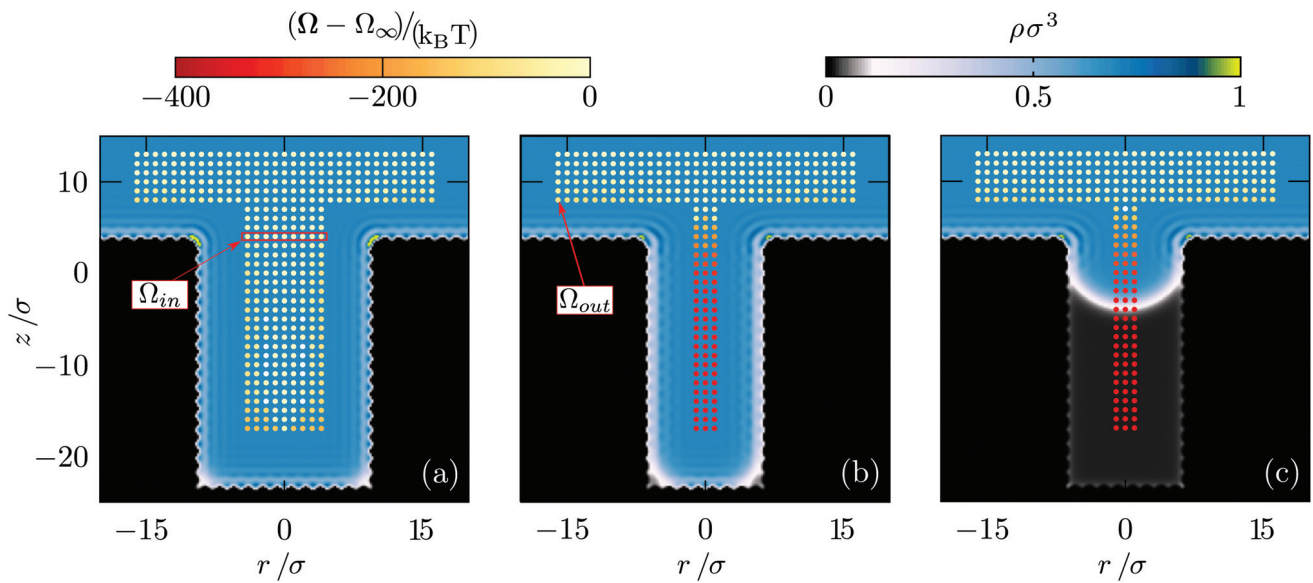

**Fig. 6** Liquid density profiles and free energy for a cylindrical pore at $\varepsilon^* = 0.1$ (a) $D_{cyl} = 20.0\sigma$, (b) $D_{cyl} = 12.5\sigma$ for a filled pore and (c) $D_{cyl} = 12.5\sigma$ for an empty pore. The color coding measures the liquid density without nanoparticle. Symbols represents the free energy at different positions of the nanoparticle near the pore. Each image is of size $40\sigma \times 40\sigma$. $\Omega_{in}$ is computed at the designated position ($r = D_{cyl}/2 + 10\sigma$) and $\Omega_{out}$ is obtained as the minimum in energy within the rectangular region.

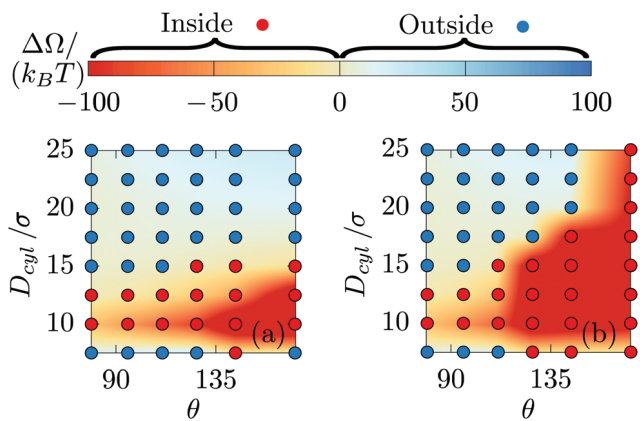

**Fig. 7** Nanoparticle entrance phase diagram at (a) $\Delta\mu = 0.27k_BT$ and (b) $\Delta\mu = 0.10k_BT$ as a function of the pore diameter $D_{cyl}$ and its hydrophobicity characterized by the contact angle $\theta$. The color coding measures the difference $\Delta\Omega \equiv \Omega_{in} - \Omega_{out}$. Red and blue points correspond to configurations where the nanoparticle prefers to stay respectively inside and outside the pore.

In an electrolyte solution, electrostatic interactions can be approximated by a simplified screened Coulomb interaction:

$$V(r)/k_BT = Z_{eff}\frac{\lambda_B}{r}\exp(-r/\lambda_D) \qquad (7)$$

where $\lambda_{B,D}$ designates respectively the Bjerrum and the Debye lengths and $Z_{eff}$ is the effective charge number.[64] The nanoparticle is thus represented as a point charge and atoms of the wall also contribute as point charges. Following the typical effective size of water molecule, we chose $\sigma = 0.275$ nm which leads to $\lambda_B = 2.5\sigma$ using for the relative water dielectric constant $\varepsilon_r = 80$.[65] The electrolyte concentration is set at $10^{-1}$ mol L$^{-1}$ which represents a reasonably concentrated ionic solution. The Debye length is then $\lambda_D = 3.5\sigma$. The obtained potential is truncated and shifted. The cutoff distance is chosen somewhat arbitrarily as the value for which the potential is reduced by a factor of 200 with respect to the screened Coulomb interaction

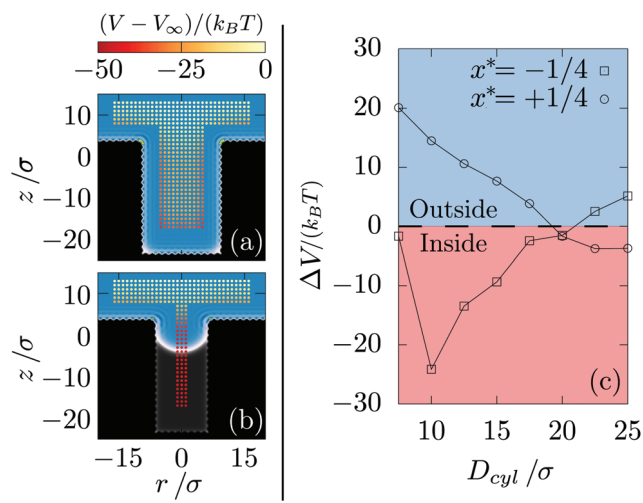

**Fig. 8** Electrostatic energy and liquid density profiles for a cylindrical pore at $x^* = -1/4$ with (a) $D_{cyl} = 20.0\sigma$ and (b) $D_{cyl} = 12.5\sigma$. The color coding measures the liquid density without nanoparticle. Symbols represents the electrostatic energy at different positions of the nanoparticle near the pore. Each image is of size $40\sigma \times 40\sigma$. (c) Difference in electrostatic energy between having the nanoparticle inside and outside for two values of $x^*$.

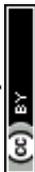

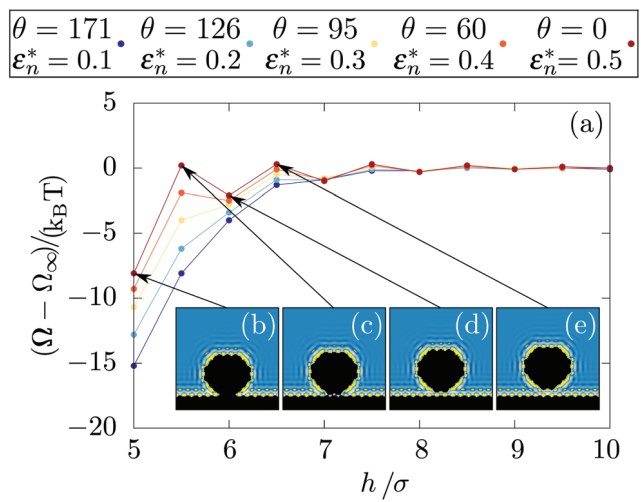

**Fig. 9** (a) Depletion potential (excess grand potential) for a nanoparticle of $R_n = 4\sigma$ as a function of the distance between the nanoparticle and a flat wall with $\varepsilon^* \equiv \varepsilon_{sol/liq}/\varepsilon_{liq/liq} = 0.3$ and different values of $\varepsilon_n^* \equiv \varepsilon_{nano/liq}/\varepsilon_{liq/liq}$. (b–e) Liquid density profiles for a hydrophilic nanoparticle near a flat wall with $\varepsilon^* = 0.3$ and $\varepsilon_n^* = 0.5$ obtained at different distances.

at $r = 1\sigma$ thus leading to a cutoff distance equal to $12\sigma$. Finally, the Coulomb interaction strength is calibrated by the effective charge number. In practice, to compare the Coulomb interaction with the depletion energy contributions, we denote by $x^*$ the ratio between the Coulomb interaction $5\sigma$ away from an infinite wall and the depletion energy for a nanoparticle of size $R = 5\sigma$ touching the most hydrophobic wall (which is equal to $-90k_BT$ [see Fig. 3]). $x^* > 0$ and $x^* < 0$ correspond respectively to repulsive and attractive Coulomb interactions and, for reference, $x^* = 1$ is equivalent to the screened Coulomb force for an effective charge, $Z_{eff} = 7.75e$. In Fig. 8(a and b), the electrostatic energy is shown at different nanoparticle positions. As with depletion energy, the minimum is located at the center of the pore below the wall. Above it, depending on the pore diameter, the minimum can be either at the pore entrance or near the flat wall [see Fig. 8c].

The calculated electrostatic contributions are added to the already obtained depletion energy. The phase diagram for nanoparticle inclusion can thus be modified for different interaction strengths [see Fig. 10]. First, adding attractive electrostatic energy allows for more conditions at which the nanoparticle can enter the pore. Then, repulsive electrostatic energy competes with the attractive depletion energy thus leading to more complex behaviors [see Fig. 10(e–h)]. Firstly, for small pores ($D_{cyl} \leq 15\sigma$), adding a repulsive electrostatic energy simply rejects the nanoparticle. Secondly, for larger pores ($17.5\sigma \leq D_{cyl} \leq 22.5\sigma$), the repulsive contribution is higher at the flat wall that at the pore entrance. Therefore, the total free energy can be located at the pore entrance which should lead to the nanoparticle entrance. Yet, the repulsive contribution is so high that the pore becomes repulsive and the nanoparticle is rejected from both the flat wall and the pore. Thirdly, for the largest pores ($D_{cyl} = 25\sigma$) there is the emergence of a region for which the nanoparticle is neither attracted nor rejected. Indeed, while the overall energy are repulsive everywhere else, at the center ($r = 0$), the nanoparticle is too far from the wall to notice its influence and there is no forces.

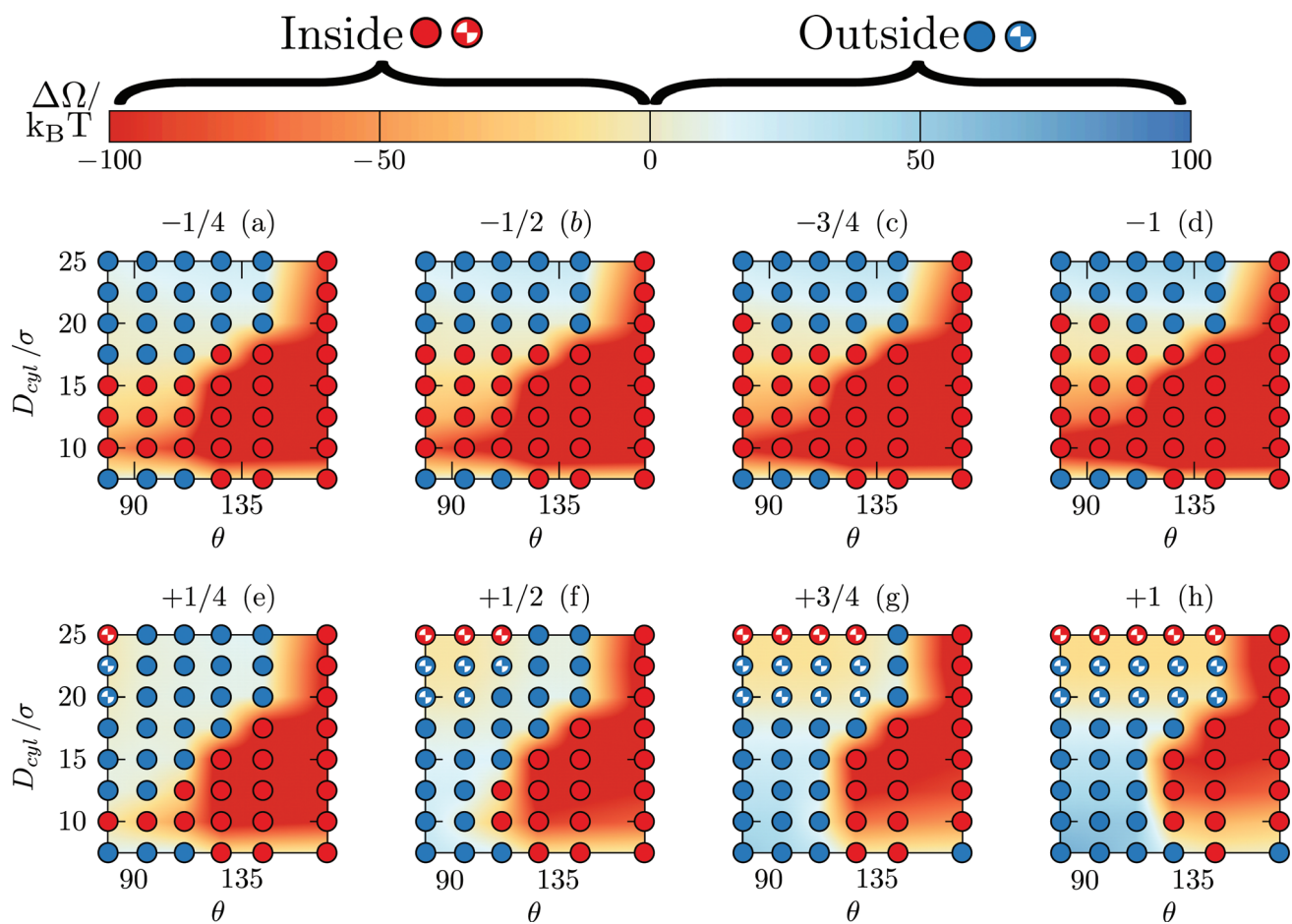

**Fig. 10** Nanoparticle entrance phase diagram at $\Delta\mu = 0.10k_BT$ when considering depletion forces and electrostatic effects for various values of $x^*$ written on top of each graphs. The color coding measures the difference $\Delta\Omega \equiv \Omega_{in} - \Omega_{out}$. (i) Filled red, (ii) filled blue, (iii) half-filled blue and (iv) half-filled black points correspond to configurations where respectively the nanoparticle prefers to stay (i) inside and (ii) outside the pore, (iii) where even if the nanoparticle prefers to stay at the pore entrance, the pore is overall repulsive and where the nanoparticle is neither attracted nor rejected from the pore.

### 3.4 Role of the nanoparticle hydrophobicity

For this final section, the nanoparticle is represented by 317 fixed particles located within a distance $R_{nano} = 4\sigma$ disposed in an FCC lattice with a solid density $\rho = 1.09\sigma^{-3}$. The truncated and shifted Lennard-Jones potential is also used to model nanoparticle–liquid interactions with the length scale and the energy parameters respectively denoted $\sigma$ and $\varepsilon_{nano/liq}$. We define $\varepsilon_n^* = \varepsilon_{nano/liq}/\varepsilon_{liq/liq}$. Fig. 9a shows the results obtained for a flat wall. When both systems are hydrophobic, the results are qualitatively similar to those obtained with a hard sphere nanoparticle [see also ESI, section 7†]. However, for hydrophilic nanoparticles, the depletion potential shows oscillations when approaching the wall. As evidenced in Fig. 9(b–e), in these cases, the nanoparticle is microscopically structured because of the large hydrophilicity and its atomistic nature. Therefore, the observed oscillations originate from the confrontation between the structures near the wall and near the nanoparticle. Then, complexity should arise when changing the wall hydrophobicity and the orientation, crystal structure and size of the nanoparticle. Those effects will be the subject of a following article.

## 4 Conclusions

In summary, our work determined the depletion potential acting on a nanoparticle near a nanoscopic pore. The size and the surface properties were systematically varied. We performed fully three-dimensional DFT calculations to determine the equilibrium structure and free energy of the system for various positions of the nanoparticle and from this extracted information about the forces driving the particle into or away from the pore. The role of functionalization of the surfaces was also considered.

We studied very specific features: (i) atomistic walls were used and crystalline patterning of the fluid near the wall was found for the first time in DFT and (ii) entrance edges playing a major role in the nanoparticle behavior. Depletion potentials were first computed for a flat wall and an infinite cylinder. We showed that the degree of hydrophobicity, the nanoparticle size but also the cylinder diameter can all increase the depletion potential. We constructed a well-defined thermodynamic model based on the capillary approximation. Reasonable agreement between the model and the DFT results shows that the model captures the main physical processes generating such a solvent mediated force. Consequently, for a coarse grain modeling, three main ingredients should be included: (i) the depletion energy is nearly linear with the distance, (ii) its onset is abrupt because it results from the emergence of a gaseous phase and (iii) its range is not larger than $5\sigma$.

Next, the nanopore geometry was studied. The first complexity arises from the fact that the liquid state is not necessarily stable inside the pore channel. A first phase diagram allowed us to identify conditions for which the pore is filled with liquid. Thereafter, a nanoparticle is positioned in the vicinity of the pore entrance. When the pore is filled, results for the depletion potential are consistent with the more simple geometries previously discussed. When the pore is empty, the depletion forces are surprisingly larger. This results from an increase of the number of particles in the liquid state when moving the nanoparticle inside the gaseous phase. Finally, a phase diagram showing conditions driving the nanoparticle inside the pore was obtained. Remarkably, while it could be thought that large pores are necessary to carry the nanoparticle inside, we demonstrate that the contrary happens. Increasing hydrophobicity and decreasing the size of the pore play two major roles. On the one hand, it increases the direct depletion forces. On the other hand, it allowed for an empty pore to be stable thus generating an additional osmotic pressure. In the last section, electrostatic effects are added by means of a simple screened Coulomb interaction. We demonstrate that tuning the amount of charges, one can trigger the nanoparticle entrance even when depletion forces prescribe otherwise.

Our microscopic models incorporate generic physical effects such as excluded volume and long (but finite) ranged attraction. This is enough to give the usual phenomenology of fluid layering near a wall, hydrophobic–hydrophilic behavior, Young's law [see Fig. 2 in ESI, section 1†], a typical liquid–vapor phase diagram, *etc.* The mechanism behind the dominant solvent-mediated forces – namely, the formation of a meniscus of vapor for hydrophobic surfaces – is also quite generic and so we expect the general result (that the solvent mediated forces have a range of about 5 solvent molecule diameters and that the energy varies linearly (and so that the forces are constant) as a function of distance from the surface) to hold for a wide variety of solvents. The only unknown quantity is, *e.g.*, the value of the binding energy at zero separation. From our analytic model for flat wall, this value can be obtained by solving a simple set of algebraic equations [see eqn (10)–(12) in ESI, section 3†].

As a perspective, the numerical values and also the thermodynamic model obtained in this work describe quantitatively the role of solvent-mediated forces. Therefore, they can be used to avoid the modeling of liquid when studying a multi-scale system made of solvent/solute/pore. In the context of selective crystallization, our work also provides an important guideline for the rational design of porous materials. Especially, we show that depletion forces can drive the nanoparticle into the pore by tuning the solvent interactions and the size of the pore. As a result, one can avoid the construction of nanoparticle specific materials. Moreover, the use of electrostatic effects allows one to adjust the nanoparticle behavior. This can happen in particular by applying an electric field to generate surface charges on the pore.

## Conflicts of interest

There are no conflicts to declare.

## Acknowledgements

The work of JL was funded by the European Union's Horizon 2020 research and innovation program within the AMECRYS project under grant agreement no. 712965. That of JFL was funded by the European Space Agency under Contract No. ESA AO-2004-070.